\documentclass[twocolumn,
               superscriptaddress,
               floatfix,
               longbibliography, 
               showkeys,apl]{revtex4}  
  
\usepackage{graphicx} 
\usepackage{bm}	
\usepackage{color}                     
\usepackage{xcolor}
\usepackage{epsfig}
\usepackage{amsmath} 
\usepackage{amssymb} 
\usepackage{longtable} 
\usepackage{float}
\usepackage{mathtools}
\usepackage{xparse}
\usepackage{hyperref}
\usepackage{times}
\usepackage[normalem]{ulem}

\usepackage{sidecap}
\sidecaptionvpos{figure}{t}

\newcommand{\be}{\begin{equation}}
\newcommand{\ee}{\end{equation}}
\newcommand{\bd}{\begin{displaymath}}
\newcommand{\ed}{\end{displaymath}}
\newcommand{\BE}{\begin{eqnarray}}
\newcommand{\EE}{\end{eqnarray}}

\definecolor{darkgreen}{rgb}{0.0, 0.5, 0.0}

\NewDocumentCommand{\ceil}{s O{} m}{
  \IfBooleanTF{#1} 
    {\left\lceil#3\right\rceil} 
    {#2\lceil#3#2\rceil} 
}

\begin{document}

\title{Classical versus Quantum Models in Machine Learning: Insights from a Finance Application}

\author{Javier Alcazar}
\affiliation{Zapata Computing Canada Inc., 1 Yonge Street, Suite 900, Toronto, ON, M5E 1E5}
\affiliation{National Australia Bank, 88 Wood St, Barbican, London EC2V 7QQ, UK}

\author{Vicente Leyton-Ortega}
\affiliation{Computer Science and Engineering Division, Oak Ridge National Laboratory,
One Bethel Valley Road, Oak Ridge, TN 37831 USA}
\affiliation{Rigetti Computing, 2919 Seventh Street, Berkeley, CA 94710-2704, USA}

\author{Alejandro Perdomo-Ortiz}
\email{alejandro@zapatacomputing.com}
\affiliation{Zapata Computing Canada Inc., 1 Yonge Street, Suite 900, Toronto, ON, M5E 1E5}
\affiliation{Department of Computer Science, University College London, WC1E 6BT London, UK}
\affiliation{Rigetti Computing, 2919 Seventh Street, Berkeley, CA 94710-2704, USA}
%end authors

\date{\today} 

\begin{abstract}
Although several models have been proposed towards assisting machine learning (ML) tasks with quantum computers, a direct comparison of the expressive power and efficiency of classical versus quantum models for datasets originating from real-world applications is one of the key milestones towards a quantum ready era. Here, we take a first step towards addressing this challenge by performing a comparison of the widely used classical ML models known as restricted Boltzmann machines (RBMs), against a recently proposed quantum model, now known as quantum circuit Born machines (QCBMs). Both models address the same hard tasks in unsupervised generative modeling, with QCBMs exploiting the probabilistic nature of quantum mechanics and a candidate for near-term quantum computers, as experimentally demonstrated in three different quantum hardware architectures to date. To address the question of the performance of the quantum model on real-world classical data sets, we construct scenarios from a probabilistic version out of the well-known portfolio optimization problem in finance, by using time-series pricing data from asset subsets of the S\&P500 stock market index. It is remarkable to find that, under the same number of resources  in terms of parameters for both classical and quantum models, the quantum models seem to have superior performance on typical instances when compared with the canonical training of the RBMs. Our simulations are grounded on a hardware efficient realization of the QCBMs on ion-trap quantum computers, by using their native gate sets, and therefore readily implementable in near-term quantum devices.
\end{abstract}
%end of abstract

\maketitle

\section{Introduction}\label{s:intro}
\newcommand{\comment}[1]{}

In the past decade, a significant interest in quantum computing has been devoted to the search of key real-world applications where quantum computers can offer a significant advantage over their classical counterparts. As in many computer science research areas, such as the development of heuristic algorithms, we expect these developments to involve experimental testing of the performance on relevant data sets, but without access to rigorous proofs or general claims about their speed up compared to previous algorithms. The development of machine learning, as applied to the real-world applications, relies largely on this approach, where new algorithms are tested and compared against other proposals via established benchmarks. 

Among the span of applications, probabilistic graphical models, and more specifically, unsupervised generative modeling stands out as one of the most promising ML areas towards a demonstration of quantum advantage with near-term quantum devices~\cite{PerdomoOrtiz2017}. It is in this domain that we focus the comparison in this work. 

To generate the benchmarks, we use stock market data and use subsets to construct a probabilistic version of a canonical problem in finance: portfolio optimization. In particular, the market data used for this study correspond to the historical time-series of stocks in index S\&P 500 for the time period encompassing daily asset pricing data between 2017-12-01 to 2018-02-07. The benchmark is constructed such that, both classical and quantum models can be compared on the same footing.

For our quantum approach, we implement here a recently proposed model referred to as quantum circuit Born machines (QCBMs)~\cite{Benedetti2019}. This model is used to load and represent arbitrary probability distributions by using the Born amplitudes of the wave function at the end of the quantum circuit, hence its name. The term Born machine was originally coined in the context of quantum wavefunctions from tensor networks objects~\cite{han2017unsupervised}. To distinguish from such quantum objects we refer to the the latter as tensor networks Born machines (TNBMs). 

QCBMs can be trained with the so-called data-driven quantum circuit learning (DDQCL) algorithm within the framework of parametrized quantum circuits and hardware efficient representations, with the help of gradient-free~\cite{Benedetti2019} or gradient-based~\cite{liu2018differentiable,Coyle2019} optimizers, making it amenable for implementations in near-term hardware. To date, experimental implementations of QCBMs via DDQCL have been implemented in ion-trap~\cite{Zhu2018} and superconducting devices~\cite{Hamilton2018,leyton2019robust}. Another recent experimental implementation was demonstrated in Ref.~\cite{Zoufal2019}, in the context of probability distributions appearing in financial applications, and where the QCBM is embedded as the generator inside a generative adversarial network approach.

As the canonical ML model to compete with the QCBMs, we choose a generative model known as restricted Boltzmann machines (RBMs). This model has been the baseline used elsewhere to compare the performance of other quantum thermal models~\cite{Amin2016, Yudong2017, Anschuetz2019}. RBMs are energy-based models that associate scalar energy to each configuration of the variables of interest. These energy-based probabilistic models define a probability distribution through an energy function analogous to Boltzmann distribution. RBMs are shallow, two-layer neural networks, where the first layer is called the visible, or input, layer, and the second is the hidden layer. Learning corresponds to modifying the parameters on that energy function so that its shape reaches configurations with low energy.

Although there have been several contributions in the frontier of finance and quantum computation~\cite{Rosenberg2016,Marzec2016, Rebentrost2018, Orus2018, Woerner2019, ORUS2019100028, Stamatopoulos2019, Yongcheng2019, Martin2019, Venturelli2019, Zoufal2019}, our focus here is in a systematic benchmark comparison of a widely known classical machine learning model with a quantum model readily implementable in near-term quantum hardware.

Without any claim to a precisely developed theory regarding the true nature of market distribution, the inspiring motivation for modeling it with the Boltzmann distribution rests on an analogy with statistical mechanics. For instance, following \cite{Matoss2011}, based on the context of an ideal gas, the companies shares negotiated by investors could be compared with particles, and therefore, under this assumption, they could be described by Boltzmann distributions. We construct such distributions from a time window of the stock market pricing data, and use them as the target distributions from which we draw the samples to be used as training sets for both competing models: the RBMs and the QCBMs. Here, the temperature can be analogously interpreted as the market temperature, capturing the variability (volatility) of the market, in analogy with an ideal gas where temperature accounts for the average agitation energy of the underlying particles. Thus, higher temperature leads to higher volatility and lower temperature of the economic system corresponds to lower volatility of the market. More details related to the construction of the benchmarks are provided in Sec.~\ref{s:problem}.

\begin{figure}[h]
\includegraphics[width=.50\textwidth]{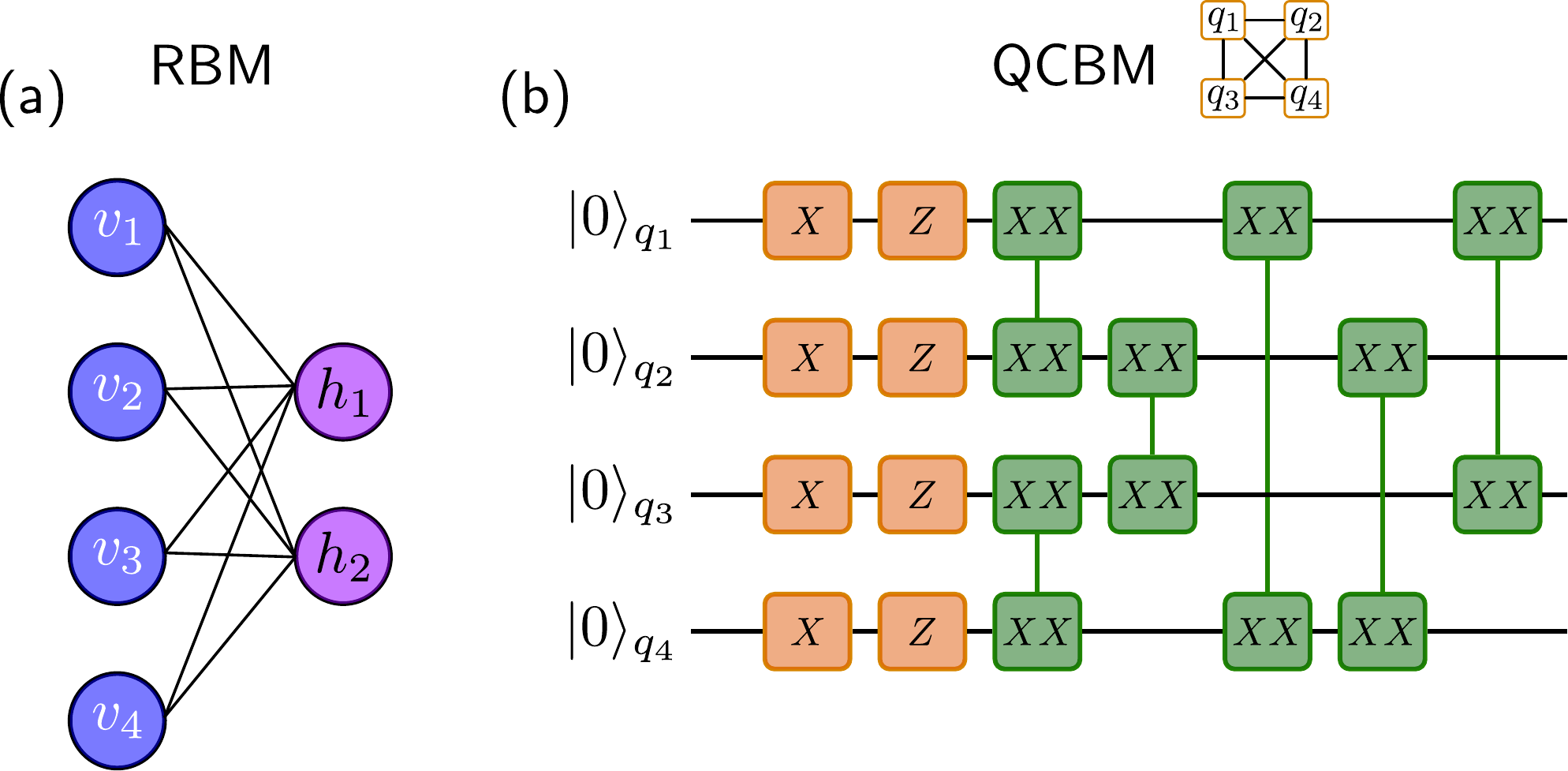}
\caption{
    {\it QCBM and RBM schematic setup}: We show an example for a subset of $N=4$ stock market assets, modeled either by the four visible nodes in the classical scheme,  $\{v_1, v_2, v_3, v_4 \}$, or by the four qubits in the quantum model. In panel (a), we show the graph layout used for the RBM training, with a number of hidden nodes set to $M=N/2$. In panel (b), we show the quantum circuit layout used for our QCBM models, with the first layer performing arbitrary single-qubit rotations through the sequence of $X$ and $Z$ rotation gates and the second layer composed of a fully connected graph of parametrized M\o lmer-S\o rensen $X\!X$ entangling gates (see Sec.~\ref{ss:qcbms} for details). This layout, inspired by the native gates in ion-trap quantum computers, contains an adjustable parameter controlling either the degree of rotation or the degree of entanglement for both single or two-qubit gates respectively. Note both schemes are designed to have the same amount of adjustable parameters to be learned in the training process: $2N + N(N-1)/2$ for QCBMs and $N+M+NM$ coming from the biases on each node and the graph weights (edges) for the case of the RBMs. Both are equal to $N(N+3)/2$. \label{f:QCBM-RBM}}
\end{figure}

As shown in Fig.~\ref{f:QCBM-RBM}, key and unique to our contribution, is that the number of parameters in both our QCBMs and RBMs is the same as the number of assets increases, therefore allowing for a fair comparison in terms of model expressivities.

In the next section, we describe the details of the benchmark proposed. In Section~\ref{s:methods} we provide the details for each of the computational approaches, and in Section~\ref{s:results} we discuss the main findings of the quantum versus classical model comparison. In Sec.~\ref{s:outlook} we summarize and point out potential research directions from this work.
%end intro

%\input{sections/problem}

\section{Problem Description}\label{s:problem}

In this section, we describe how to transform data taken from the stock market into a probabilistic model that can be used to generate the training set for both, the quantum and classical models (QCBMs and RBMs, respectively).

The selection of optimal investment portfolios is a problem of great interest in the area of quantitative finance. The problem is of practical importance for investors, whose objective is to allocate capital optimally among assets while respecting some investment restrictions. The goal of this optimization task, introduced by Markowitz~\cite{Markowitz52}, is to generate a set of portfolios that offer either the highest expected return (profit) for a defined level of risk (standard deviation of portfolio returns) or the lowest risk for a given level of expected return. This set represents the so called {\it efficient frontier} (EF).

More precisely, the portfolio optimization model aims at determining the fractions $w_i$ (such that $\sum_i^N w_i = 1$) of a given capital to be invested in each asset $i$ of a universe of $N$ assets, such that these minimize the risk $\sigma (\boldsymbol{w})$ for a given level $\rho$ of the expected return $\langle r(\boldsymbol{w}) \rangle$. The problem can be written as:
\begin{eqnarray} \label{eq:optProblem}
{\rm min} \left\lbrace \sigma^2(\boldsymbol{w})   =  \sum_{i = 1}^N \sum_{j = 1}^N \right.
\Sigma_{ij} w_i   w_j &:& \nonumber \\    \langle r (\boldsymbol{w}) \rangle &=& \left.
\sum_{i=1}^N \mu_i w_i = \rho \right \rbrace , \nonumber \\ 
\end{eqnarray}
where $\Sigma_{ij}$ is the sample covariance obtained from the return time series of asset $i$ and asset $j$, and $\mu_i$ is the average of the return time series of asset $i$, with each daily return, $\mu_i^t$, calculated as the relative increment in asset price from its previous day (i.e., $\mu_i^t = (p_i^t - p_i^{(t-1)}) / p_i^{(t-1)}$, with $p_i^t$ as the price for asset $i$ at time $t$). We denote by $\sigma_\rho$ the outcome from Eq.~\ref{eq:optProblem} for a given return level $\rho$.  The graph $\varphi$ of the pairwise $( \sigma_\rho , \rho)$, for different values of $\rho$ on a given interval $[\rho_0, \rho_1]$, coincides with the set of all efficient portfolios (i.e., the {\it efficient frontier}). Under no further constraints, solving for $\sigma_p$ from Eq.~\ref{eq:optProblem} can be done efficiently with quadratic programming (QP) algorithms.

Note the optimization task in Eq.~\ref{eq:optProblem} has the potential outcome of investing small amounts in a large number of assets, as an attempt to reduce the overall risk by ``over diversifying" the portfolio. This type of investment strategy can be challenging to implement in practice: portfolios composed of a large number of assets are difficult to manage and may incur in high transaction costs.  Therefore, several restrictions are usually imposed on the allocation of capital among assets, as a consequence of market rules and conditions for investment or to reflect investor profiles and preferences. For instance, constraints can be included to control the amount of desired diversification, i.e., modifying bound limits $\lbrace l_i, u_i \rbrace$ to the proportion of capital invested in the investment on individual assets or a group of assets, thus the constraint $l_i < w_i < u_i$ could be considered.

A more realistic and common scenario is to include in the optimization task a {\it cardinality constraint}, which limits directly the number of assets to be transacted to an pre-specified number $\kappa < N$. Therefore, the number of different sets to be treated is $M = \binom{N}{\kappa}$. In this scenario, the problem can be formulated as a Mixed-Integer Quadratic Program (MIQP) with the addition of binary variables $x_i \in \{0,1 \}$, for $i = 1,..., N$, which are set to ``1" when the $i$-th asset is included as part of the $\kappa$ assets, or ``0" if it is left out of this selected set. The optimization task can be described as follows:

\begin{eqnarray} \label{eq:KoptProblem}
  {\rm min} \left \lbrace  \sigma^2(\boldsymbol{w}) \right.&& : \nonumber \\
  && \langle r(\boldsymbol{w}) \rangle = \rho , \nonumber \\   
  && l_i x_i < w_i < u_i x_i \quad i=1,...,N, \nonumber \\
  && \sum_{i=1}^N x_i = \left. \kappa \right\rbrace   .   
\end{eqnarray}

\comment{In this problem we denote by $\sigma_\rho^\kappa$ the outcome from \ref{eq:KoptProblem} for a given return level $\rho$. The graph $\varphi^\kappa$ of pairwise $( \sigma_\rho^\kappa , \rho)$, i.e. the efficient frontier, is no longer convex neither continuous in contrast with $\varphi$ defined in problem \eqref{eq:optProblem}.}

In general terms, a possible approach to solving the optimization problem \eqref{eq:KoptProblem} is to enumerate all possible subsets of $\kappa$ assets and, for each of them, to solve the associated QP that considers only the assets in the subset for the optimization. However, such an exhaustive enumeration scheme is not practical in this context, as this MIQP problem falls in the class of considerably difficult NP-hard problems~\cite{MoralRuiz2006}. As this brute-force approach appears infeasible, other heuristic avenues have been proposed to tackle the hybrid optimization problem, concurring on a strategy consisting of breaking the problem on a continuous part solved via QP and leaving the discrete to be dealt with kind of black-box solvers, as genetic algorithms, particle swarm optimisation or dimensionality reduction (see e.g., Refs.~\cite{Kresta11, Farzi13, Rifki12, Ruiz10}). Alternatively to heuristic methods, this problem may be suitable for quantum computers, which can help in the discrete hard part of the problem, resulting in a hybrid combination of classical and quantum resources for the continuous and discrete parts, respectively. We describe next a probabilistic variant from this hard optimization problem that will allow us to compare the performance of RBMs versus QCBMs).

As indicated above, the different potential portfolios with $\kappa$ assets are encoded in bitstrings of size $N$ (represented by the variable $\boldsymbol{x}$ in Eq.~\ref{eq:KoptProblem}). Therefore,  every bit represents the inclusion of that asset in the candidate portfolio ( $1$ - selected and $0$ - not selected), and the valid portfolios would have a number $\kappa$ of $1$'s, as specified in the cardinality constraint. For instance, for $N=4$ and $\kappa = 2$, the different valid configurations can be encoded as $ \{ 0011, 0101, 0110, 1001,1010,1100 \}$.

\begin{figure}[h!]
  \includegraphics[width=.45\textwidth]{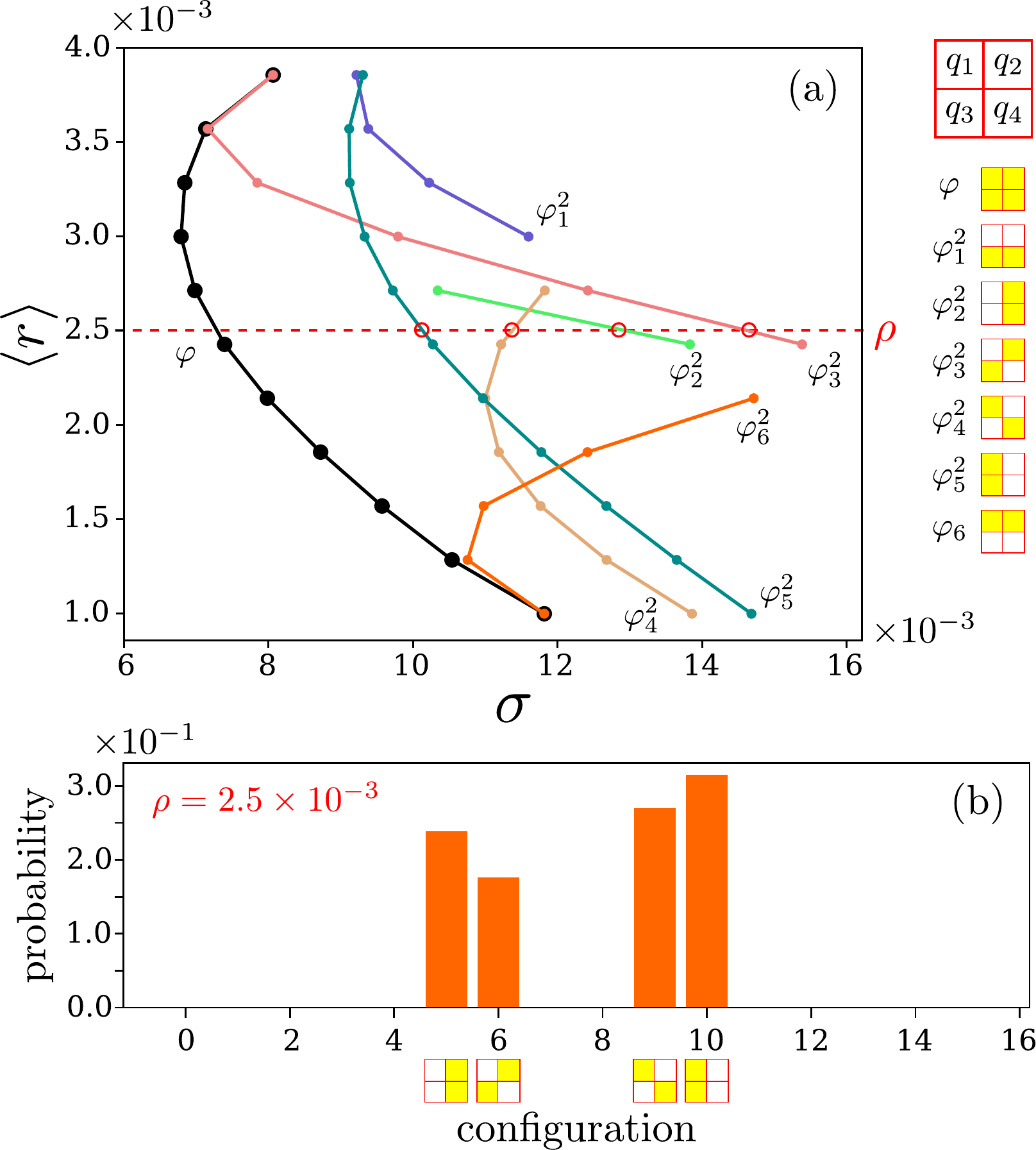}  
  \caption{{\it Probabilistic model from portfolio optimization problem}: Panel (a) shows the efficient frontiers corresponding to the solution of a 4-asset subset taken from the S\&P 500 stock index under the unconstrained case (Eq.~\ref{eq:optProblem}, and labeled as $\varphi$) and from the constrained optimization problem (Eq.~\ref{eq:KoptProblem}, labeled as $\varphi_i^{2}$ for $i = 1, ... , 6$), where two assets can be selected at a time ($\kappa = 2$). Note that the optimal risk for the unconstrained case can differ significantly from the constrained case, as can be seen in the example for the expected return, $\rho =   2.5 \times 10^{-3}$. Optimal values correspond to $x$-axis values corresponding to the crossing (red circles) of the fixed $\rho$ line and the efficient frontiers curves, obtained from solving each QP problem. From each of these solutions (crossings) we can associate a Boltzmann probability according to Eq.~\ref{boltzdist}, where the portfolios with lowest risk would be sampled with more probability. In panel (b) we show the probability distribution resulting from the example in (a) at $\rho =   2.5 \times 10^{-3}$. This would correspond to one of the many target distributions used as benchmarks and to be learned by both the RBM model and the QCBM model. For illustration, we encode the different assets configuration by 4-pixel binary images [see encoding in panel (a)], where full pixels indicated the assets selected. For example, the configuration $1001$ (\#9 in the $x$-axis) corresponds to the only asset \#1 and \#4 being selected.
  }
  \label{f:EF}
\end{figure}

The MIQP problem in Eq.~\eqref{eq:KoptProblem} is tantamount to multiple QP problems over each possible set of $\kappa$ assets defined by the cardinality constraint. For instance, in the aforementioned case with $N=4$ and $\kappa=2$, the problem is equivalent to $M=6$ different QP problems. The solution for each of those QP problems yields the efficient frontiers $\varphi_{i}^\kappa$, for $i =1, ..., M$, which are enveloped by the efficient frontier of the optimization task without the cardinality constraint, as illustrated in Fig.~\ref{f:EF}(a). The efficient frontier $\varphi_{i}^\kappa$ encloses all efficient portfolios for the $i$th possible sets or configurations of $\kappa$
assets.  

The next step in completing the genesis of the training data set corresponds to generating a probabilistic distribution out of the different aforementioned efficient frontiers. To do so, we determine the risk $\sigma_{\rho,i}^\kappa$ for each of the $M$ efficient frontiers for a given return level $\rho$,  and assign the probability of finding this configuration by the Boltzmann distribution
\begin{eqnarray}
&p(\sigma_{\rho,i}^\kappa) = \dfrac{ e^{ -\sigma_{\rho,i}^\kappa / T }
}{\sum_{i=1}^M e^{ - \sigma_{\rho,i}^\kappa / T} },
\label{boltzdist}
\end{eqnarray}
where $T$ is referred to as the market temperature. This temperature parameter stands as an analogous representation of volatility in the market, where higher temperature means higher volatility and lower temperature of the economic system is interpreted as lower volatility of the market. With these considerations, we define the temperature parameter in Eq.~\ref{boltzdist} as the square root of the mean of the covariance matrix defined in Eq.~\ref{eq:optProblem}, as this matrix encapsulates the risk information (volatility) as stated in the Markowitz's model.

With this encoding, we are able to represent the different probabilities (given by Eq.~\eqref{boltzdist}), and from which we can draw samples that would define the target dataset $\cal D$ to be learned by the QCBMs and the RBMs. In Fig.~\ref{f:EF}(b) we plot the probability of every configuration,  where $\{ 0101,0110,1001,1010\}$ are the configurations with efficient portfolios for a expected return $\rho = 2.5 \times 10^{-3}$ (red dotted line in Fig.~\ref{f:EF}(a)). Note for two out of the six possible configurations there is no solution to the QP problem under the specified value of $\rho$, since there is no crossing to the red dotted line for $\varphi_1^2$ and $\varphi_6^2$. For those configurations, a risk of effectively infinity is assigned such that probability is zero. This is the reason why Fig.~\ref{f:EF}(b) shows only four peaks, instead of six.   

In Sec.~\ref{s:results} we provide more details about the construction of the 30 realizations of target distributions per system size, i.e., number of stock assets. With the intention of having statistical diversity in the benchmarks, these target distributions include different values of $\rho$ and different random subsets from the S\&P 500 for each system size.

\section{Methods}~\label{s:methods}

%\vspace{-1.5cm}
\subsection{The quantum learning pipeline}~\label{ss:qcbms}

The QCBM model is the hybrid quantum-classical algorithm we consider here for the quantum learning approach since it is tailored towards generative modeling in unsupervised ML, i.e., it aims to capture the target benchmark distribution through a quantum wavefunction (for more details, see Ref.~\cite{Benedetti2019}). This algorithm uses a parametrized quantum circuit (PQC), with fixed depth and gate layout, that prepares a wavefunction $|\psi(\boldsymbol{\theta}) \rangle $  from which probabilities are obtained according to Born's rule $P_{ \boldsymbol{\theta} }(\boldsymbol{x}) = | \langle \boldsymbol{x} | \psi ( \boldsymbol{\theta} ) \rangle|^2 $. The $N$-dimensional binary vectors $\boldsymbol{x} \in \{0, 1 \} ^ N$ are associated with the so-called computational basis of the $N$-qubit quantum states, e.g. $1010 \rightarrow |1010\rangle$, and as mentioned above, these map one-to-one to valid portfolios in this $2^N$ configuration space. A classical solver updates the quantum circuit parameters $\boldsymbol{\theta}$ in pursuit to minimize the Kullback-Leibler (KL) divergence $D_{KL}(P_{\cal D} | P_{\boldsymbol{\theta}} )$. The latter measures how the circuit probability distribution (learned distribution) $P_{\boldsymbol{\theta}}$ is different from the target probability distribution $P_{\cal D}$. To evaluate this loss
we consider the Kullback-Leibler divergence ($D_{KL}$) defined as

\begin{eqnarray}
    D_{KL} [ P | Q ] = \sum_{\boldsymbol{x} \in \{ 0,1\}^N} P(\boldsymbol{x}) \ln P(\boldsymbol{x}) \nonumber \\ 
    - \sum_{\boldsymbol{x} \in \{ 0,1\}^N} P(\boldsymbol{x}) \ln ( {\rm max } ( \epsilon, Q(\boldsymbol{x}))),   
\end{eqnarray}

where we have introduced a clip $\epsilon$ to avoid singularities when $Q(\boldsymbol{x}) = 0$. 

The quantum circuit should be able to prepare a broad range of wavefunctions, in order to approach a given target distribution. To this end, we consider a general circuit parametrized by single-qubit rotations (first layer) and two-qubit entangling rotations (second layer). Inspired by the gates readily available in ion trap quantum computers, we use M\o lmer-S\o rensen $X\!X_{ij}$ entangling gates for the second layer, where $X\!X_{nm} (\chi) = \exp(-i \sigma_n^x \sigma_m^x \chi \pi/2)$, with $\sigma_n^{\alpha}$ the Pauli operators for $\alpha  = x, y$, and $z$. In this ansatz, the number of parameters depends only on the number of qubits $N$.

In the first layer, since we execute circuits always from the ground state $|0...0\rangle$, it is enough to apply single-qubit operations relying on $X_n(\theta) = \exp(-i \sigma_n^x \theta \pi/2)$ and $Z_n(\theta) = \exp(-i \sigma_n^z \theta \pi/2)$  rotations as it is shown in Fig. \ref{f:QCBM-RBM}. After, in the second layer, we can perform $X\!X_{ij}$ entangling gates involving any two qubits following a fully connected graph. Thus, under this two-layer ansatz, single-qubit operations would require $2N$ parameters and the two-qubit entangling gates would require $N(N-1)/2$ parameters, summing up to $N(N+3)/2$ circuit parameters to be learned with the help of the classical solver. 

The quantum circuit simulations are performed using Rigetti's quantum virtual machine (QVM); it is a part of the Forest\textsuperscript{TM} SDK available in \cite{forest}. Additionally, for our simulations we assume a noiseless device and infinite measurement precision allowing us to compute the Born probabilities of the quantum probabilistic model directly from the computed wavefunction amplitudes.

\subsection{The classical learning approach}

We consider the widely used RBM model as the classical counterpart of the aforementioned QCBM. We consider the RBM standard type, that consists of binary variables, classified as hidden $\{ h_1, h_2, ..., h_{N_h} \}$ and visible $\{v_1, v_2, ..., v_{N_v} \} $ units or nodes, with $N_h < N_v$. This approach generates a probabilistic model $P(\boldsymbol{h},\boldsymbol{v}) = \exp [-E(\boldsymbol{h},\boldsymbol{v})]/Z$, based on the energy $E(\boldsymbol{h},\boldsymbol{v}) = \sum_i  \theta_i^v v_i + \sum_i 
\theta_j^h h_j + \sum_{ij} 
\theta_{ij} v_i h_j$. The normalization term $Z$ is the partition function. The energy $E(\boldsymbol{h},\boldsymbol{v}) $ is biased by $N_v + N_h $ weights (one per node), and $N_h \times N_v$ weights associated with the connections between hidden and visible nodes. These weights play the role of parameters to be learned in the learning process. We consider $N_h = N_v/2$ and $N_v = N$, for $N$ even, to get the same number of parameters used for the QCBM (see Fig. \ref{f:QCBM-RBM} for an example of a universe of four available assets). 

This learning approach stands on gradient-based maximization of the likelihood of the RBM's parameters given the training data. We consider the persistent contrastive divergence (PCD) training~\cite{Hinton2006, Tieleman2008},  in which a Gibbs chain is run for only $K_{\rm RBM}$ steps to estimate the log-likelihood gradient given the data set. For the implementation, we used Theano \cite{TheanoRBM}.
%end of methos
 
%\input{sections/results}
\section{Results and Discussion}\label{s:results}

For our simulations, we consider as our universe of stocks subsets of the historical time-series from the S\&P500 index for the period encompassing 2017-12-01 to 2018-02-07. These stocks subsets are constructed by randomly selecting $N$ assets out of the full index aforementioned. For a robust comparison, we consider several target distributions generated by the set of six expected return levels. Given the historical return values, we chose these levels of return to be $\{0.010, 0.015, 0.020, 0.025, 0.30, 0.35 \}$. In all of our numerical experiments with $N$ assets, the cardinality constraint was set to $\kappa = N/2$, and $N \in \{ 4, 6, 8 , 10\} $. Additionally, we consider five different random subsets per problem size $N$, which altogether with the six return levels, it makes for a total count of 30 different distributions to be learned by the QCBM and RBM per each $N$. 

For the quantum circuit learning algorithm, the number of qubits $N$ is equal to the number of considered stock market assets and we use the CMA-ES solver \cite{hansen2019pycma,hansen2001} for the optimization part, with variable standard deviation, to minimize $D_{KL}$. In the classical learning scheme, we consider the $K_{\rm RBM} = 1$ standard value for the Gibbs sampling in the PCD training, but higher values of $K_{\rm RBM}$ were also considered (see Appendix \ref{s:other-kvalues} for results with $K_{\rm RBM} = 10$ and $100$). Due to the stochastic nature of the starting point in the learning procedure, we consider 11 repetitions of each simulation to evaluate the learning scheme typical performance by calculating their $D_{KL}$ median value. 

We collect all the median values of $D_{KL}$ using QCBM and RBM, for a given $N$, and bootstrap those values in a sample size of $10^4$. The typical performance is given by the median of this bootstrap analysis, with a confidence interval given by the $5$th and $95$th percentile of that median. 

\begin{figure*}
\includegraphics[width=.9\textwidth]{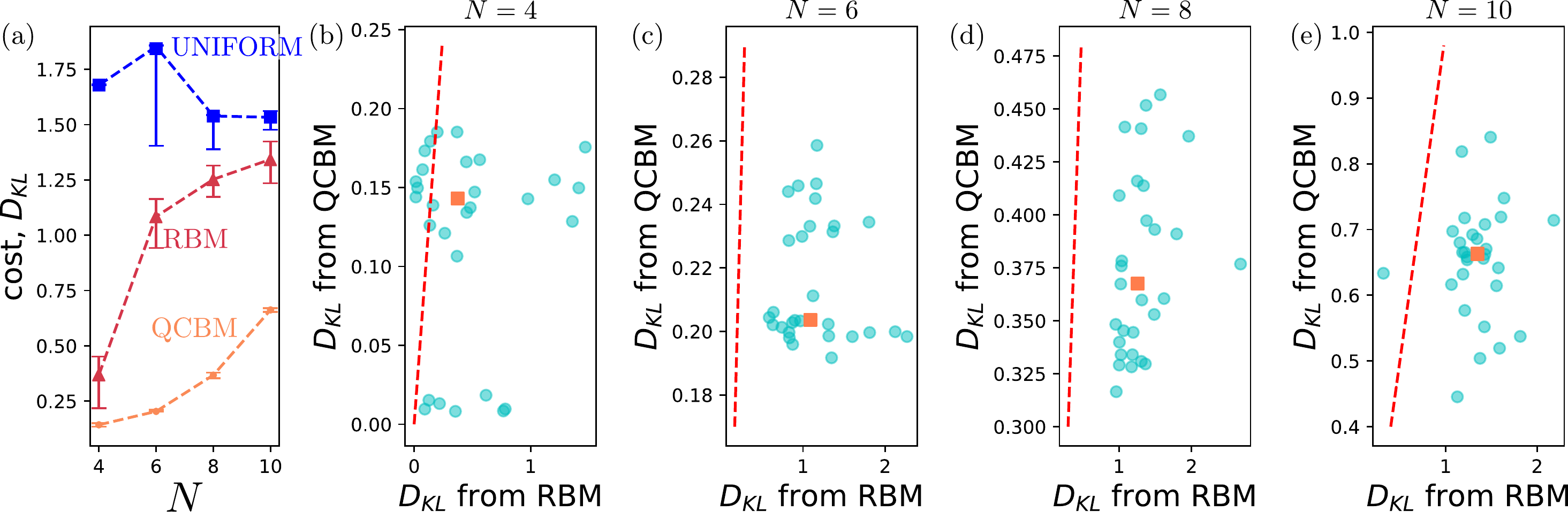}   
\caption{\label{f:scaling} 
{\it Performance scaling comparison for quantum and classical learning schemes}: In this figure, we show the score of the learned distribution using classical and quantum algorithms. In panel (a) we present a comparison between the bootstrapped median value of $D_{KL}$ from 30 different scenarios per problem size $N$ comprising of different stock market asset selection, under a cardinality constraint setting of $\kappa = N/2$ and several expected return levels. Error bars depict 5 and 95 percentile (90\% confidence interval) of each bootstrapped median value of the cost. The line corresponding to the uniform distribution over all the outcomes serves as a baseline since no training is required to generate such model. Panels (b) --(e) show scatter plots of the costs from QCBM against those cost from RBM, for $N=4,6,8,$ and $10$.  The orange square-symbol datapoint corresponds to the median value reported in (a). The red dashed line corresponds to the $D^{\rm{RBM}}_{KL} = D^{\rm{QCBM}}_{KL}$ line and therefore,  points falling bellow this line correspond to those where the performance of the QCBM model is better than the RBM model}
\end{figure*}

From Fig.~\ref{f:scaling}(a) it can be seen that the quantum model clearly outperforms the classical ML model on typical instances under the assumptions presented here.

For a better analysis of the results,  we compare the performance using a two dimensional scatter plot composed by the QCBM and RBM results. In this way, we can visualize the relative performance by the position of the points with respect to the identity line where $D^{\rm{RBM}}_{KL} = D^{\rm{QCBM}}_{KL}$ . In Figures \ref{f:scaling}(b) -- (e), we show the scatter plots for different values of $N$, in which most of the points fall below the identity line indicating the superior performance of QCBM model is not just on average, i.e., for  typical instances, but that it corresponds to a close to 100\% win when looked on a case-by-case basis. For the smallest size, $N=4$, the RBM performance can be enhanced using larger amounts of Gibbs sampling $K_{\rm RBM}$. In Appendix \ref{f:other-kvalues} we consider the oversize values $K_{\rm RBM} = 10$ and $100$, in those plots most of the points in the scatter plot falls above the identity line for $N =4$. But most importantly, for other problem sizes with $N=6,8$ and $10$, the performance of QCBM still largely outperforms RBM even on average as dictated by the median and on a case-by-case basis as well (see Fig. \ref{f:other-kvalues}).

Additionally, we consider in Fig.~\ref{f:scaling}(a) the performance of a uniform probability distribution over all the potential $N$ asset outcomes. This test helps to define upper limit values for $D_{KL}$ in the scenario where no learning or educated guess criteria is involved, and therefore to determine the usefulness of the classical or quantum model training from their corresponding values. As can be seen from the results, the training of the classical RBM tends toward that limit for the larger problem sizes, hinting that this energy model with quadratic interactions and $N/2$ hidden units does not have enough expressive power or that there are difficulties in training such models. For example, another explanation for the significant underperformance of the classical RBM results might be due to the so-called ``curse of dimensionality", where the performance is significantly affected by the increase in the number of parameters as the problem size increases. In other words, the model optimization process gets lost in parameter space. 
Given that both classical and quantum models are trained with the same number of parameters, it is very encouraging that our QCBM training still finds models with $D_{KL}$ values further away from this uniform probability distribution reference. We have chosen this probability as the reference since it is the most trivial model realization that already exists in the search space of both parametrized probability distributions, from the RBMs and the QCBMs; in the RBMs it would be equal to setting every $\theta^v_i= \theta^h_i=\theta_{ij} = 0$  and in the QCBM to just applying the equivalent of Hadamard gates in the first layer, therefore preparing a uniform superposition in the computational basis.

%end of results

%\input{sections/outlook}
\section{Outlook}\label{s:outlook}

Understanding whether quantum models designed for near-term quantum devices could have an impact in industrial-scale applications is one of the most pressing milestones towards a quantum ready era. In this work, we perform a direct comparison between QCBMs and RBMs, as two proposed models for tackling generative models in unsupervised ML. Although our results are positive news for the quantum-based models within a concrete probabilistic framework of portfolio optimization, several research directions could be addressed as future work.

A natural extension of this work would be to include quantum simulations and computational experiments to a larger number of qubits.  Ideally, for it to be of interest from a commercial point of view, we would require to reach the low hundreds of assets, and without resorting to preprocessing techniques, in our QCBM model this would amount to an equivalent number of qubits. Although not achievable \textit{in-silico} via quantum simulations, it is positive that such sizes are available within the NISQ era, assuming a steady pace of the development of current technologies. Certainly simulations with $\sim$ 40 number of qubits can be reached, and this could be close to the limit that can be reached without resorting to supercomputing centers to perform the quantum simulations of the quantum circuits and the generation of the training data set itself for our proposed benchmarks. Although the circuit ansatz used for the comparison to RBM models is readily implementable on ion-trap NISQ devices, it would be interesting to extend the comparison to native gates and PQC on other hardware architectures, such as those in quantum processors based on superconducting qubits.

Another potential extension to consider emerges from the decision related to the number of samples withdrawn from the target distribution, and used as the training set for both classical and quantum models. In this work, this number of samples was high enough to reach the limit where we had an accurate representation of the target distribution itself. The reason behind this experimental design choice was to focus on answering the question of the capabilities of QCBMs or RBMs to represent the target distributions, without binding our conclusions to another hyperparameter dictating the size of the training set. One interesting question arises from this decision and it is  part of ongoing work from this contribution. What are generalization capabilities of each model and their statistical efficiency? This is, which model would be able to better capture the features of the target distribution, given a very limited amount of observed data as part of the training set? This question will be addressed in a subsequent version of this work.

Finally, we wanted to conclude with some remarks about the reasoning behind the financial model and the design of the benchmarks. Although the probabilistic framework proposed here is not the canonical portfolio selection optimization problem, the construction provides a benchmark scheme which is not only generated from real stock market price data over a period of time but that also describes a desired target distribution which captures investment risk. It is important to note that computing such target distributions is intractable for the case of a large number of assets since it involves solving for each of the portfolio options in this combinatorial search space. Nevertheless, approaching this optimization problem with an iterative probabilistic Bayesian approach could be an interesting computational strategy in its own towards the hard constrained optimization problem. This can be achieved, for example, by starting from a handful of portfolios, evaluating their costs, and performing Bayesian updates until one finds a set of candidates with the desired investment configurations.  

Since we are not addressing the optimization problem itself directly, we prefer to refer to our approach as an \textit{application-inspired benchmark}, given that it has many elements of reality (the data itself and the Markowitz model to estimate the investment risk) but the probabilistic construction is tailored to be able to answer questions related to the complexity of quantum and classical ML models and to compare them on the same footing. It is important to mention that the benchmarking approach proposed here is not limited to the Markowitz model but can be extended to other portfolio optimization strategies based on other risk metrics. In any other model, we can associate the value of the function to be optimized as the energy term in the Boltzmann distribution, and the temperature would need to be scaled accordingly.

Having benchmarks similar to the one presented here and built on real data in contrast to synthetic data or synthetic models, could be a gateway to answer one of the key questions in the field: what are the features from classical distributions that could be more amenable to quantum, rather than to classical models? In other words, where and how can we carve for quantum advantage in real-world scenarios and in problems of commercial interest? This ``more amenable" qualification can be measured with many figures of merits. In our work, we decided to use the number of parameters allowed for both models, since this limits the number of degrees of freedom allowed to each model, but some other practical figures of merit to consider could be the time of execution, the energy consumption and/or simply the cost of computational time. Our approach consisted in addressing the comparison to the canonical and widely used model in ML known as RBMs. Although a good starting point to compare to our quantum counterpart, further exploration should include other modern ML generative models such as variational encoders (VAEs), generative adversarial networks (GANs), and quantum-inspired models such as TNBMs~\cite{han2017unsupervised,gao2017efficient,Glasser2019,Bradley2019}; we performed preliminary studies to include such cases  but an apples-to-apples comparison might be challenging given that these are usually tailored to tackle probability distributions over continuous variables and matching the number of parameters might be challenging as well. Although examples of benchmarks using real-world scenarios or data are limited (see e.g., Ref.~\cite{PerdomoOrtiz2017a}), these provide valuable and unique insights into the power of quantum models and algorithms which might differ significantly from benchmarks with synthetic ones on random generic problems. We believe these approaches would be key to the development of hardware and quantum algorithms towards a demonstration of quantum advantage on real-world applications. 
%end of outlook

%\input{sections/acknowledgements}  
\begin{acknowledgments} 

The authors would like to acknowledge Marcello Benedetti, Dax Koh, and Yudong Cao for useful feedback on an early version of this manuscript. V.L-O was supported by ASCR Quantum Testbed Pathfinder Program at Oak Ridge National
Laboratory under FWP \#ERKJ332.

\end{acknowledgments}
%end of acknowledgements

\onecolumngrid{}

\clearpage
\newpage
\appendix

\section{Results for PCD-$K_{\rm RBM}$ , with $K_{\rm RBM} >1$}\label{s:other-kvalues}

 %\begin{widetext} 
\begin{figure*}[h] 
\centering 
\includegraphics[width=.85\textwidth]{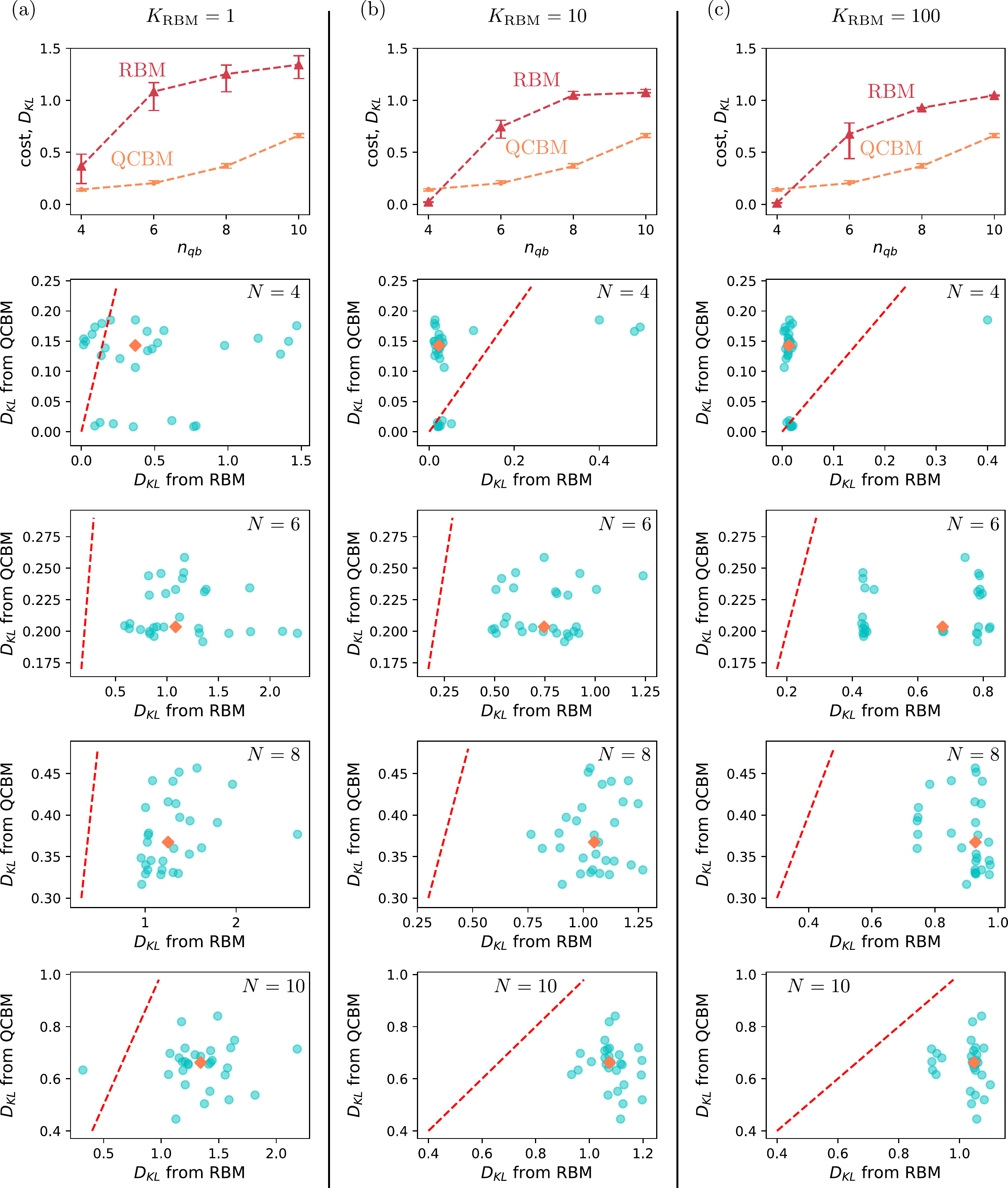} 
\caption{{\it Performance scaling for quantum and classical learning schemes for
$K_{\rm RBM} \geq 1$}: In this figure, we show three main panels with the results for (a) $K_{\rm RBM}=1$, (b) $K_{\rm RBM}=10$ and (c) $K_{\rm RBM}=10$. Every panel is headed by the comparison between the median value of $D_{KL}$ out of all assets configuration due to the cardinality constraint and all expected return levels. Below of this, we show scatter plots of the costs from QCBM and RBM for different amounts of available assets, for $N=4, 6, 8, $ and $10$. We depict 5 and 95 percentile of each median value of the cost as error bars.}\label{f:other-kvalues} 
\end{figure*} 
%\end{widetext}

%\input{sections/app}

\clearpage
\twocolumngrid{}
%\bibliography{../../bibs/refsAQCall_06202017,../../bibs/new,../../bibs/refs_katzgraber}

\end{document}